\documentclass[12pt]{iopart}

\usepackage{iopams}
\usepackage[english]{babel}
\usepackage{graphicx}
\usepackage{dsfont}
\usepackage{cite}

\begin{document}

\title[Sequential pattern formation governed by signaling gradients]{\sffamily \LARGE Sequential pattern formation governed by \\ signaling gradients}

\author{\sffamily David J. J\"org$^\mathsf{1,2,*}$, Andrew C.~Oates$^\mathsf{3,4}$, and Frank J\"ulicher$^\mathsf{1,2}$}
\address{\sffamily $^\mathsf{1}$ Max Planck Institute for the Physics of Complex Systems, N\"othnitzer Str. 38, \\ \ \, 01187 Dresden, Germany}
\address{\sffamily $^\mathsf{2}$ Center for Advancing Electronics Dresden cfAED, 01062 Dresden, Germany}
\address{\sffamily $^\mathsf{3}$ Francis Crick Institute, Mill Hill Laboratory, The Ridgeway, Mill Hill, \\ \ \, London NW7 1AA, UK}
\address{\sffamily $^\mathsf{4}$ University College London, Gower Street, London WC1E 6BT, UK}
\address{\sffamily $^\mathsf{*}$ Present address: Cavendish Laboratory, Department of Physics, \\ \ \, University of Cambridge, JJ Thomson Avenue, Cambridge CB3 0HE, UK and \\ \ \, The Wellcome Trust/Cancer Research UK Gurdon Institute, University of Cambridge, \\ \ \, Tennis Court Road, Cambridge CB2 1QN, UK}
\ead{julicher@pks.mpg.de}

\begin{abstract}
\noindent 
	Rhythmic and sequential segmentation of the embryonic body plan is a vital developmental patterning process in all vertebrate species.
	However, a theoretical framework capturing the emergence of dynamic patterns of gene expression from the interplay of cell oscillations with tissue elongation and shortening and with signaling gradients, is still missing.
	Here we show that a set of coupled genetic oscillators in an elongating tissue that is regulated by diffusing and advected signaling molecules can account for segmentation  as a self-organized patterning process.
	This system can form a finite number of segments and the dynamics of segmentation and the total number of segments formed depend strongly on kinetic parameters describing tissue elongation and signaling molecules.
	The model accounts for existing experimental perturbations to signaling gradients, and makes testable predictions about novel perturbations.
	The variety of different patterns formed in our model can account for the variability of segmentation between different animal species.
\end{abstract}


\vfill

\noindent {\footnotesize \sffamily
This is a preprint of the corresponding article published as \\
\textit{Physical Biology} \textbf{13}, 05LT03 (2016).
}

%\submitto{Physical Biology}

\maketitle

\begin{figure}[b]
\begin{center}
\includegraphics[width=10.5cm]{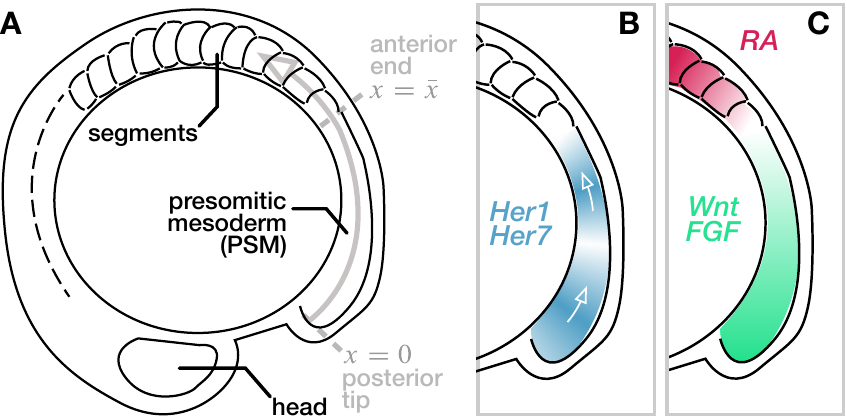}
\caption{(A) Schematic depiction of a zebrafish embryo during the segmentation of the body axis. (B) Oscillations of gene activity of cyclic genes (\textit{Her1}, \textit{Her7}) manifest themselves as traveling waves of gene expression through the PSM (blue). Arrows indicate the direction of wave propagation. (C) Concentration profiles of the signaling molecules Wnt and FGF with highest concentration at the posterior tip (green) and an opposing gradient of Retinoic Acid (RA, red) having highest concentration in the segments.}
\label{fig.1}
\end{center}
\end{figure}

\noindent%
Morphogenesis, the formation of shapes and patterns in the developing embryo, relies on the tight coordination of cellular actions \cite{Wolpert2006}.
During embryonic development, spatial profiles of signaling activity in tissues control the behavior of cells such as proliferation, migration, and differentiation \cite{Turing1952,Rogers2011,Werner2015}. 
In vertebrates, a vital morphogenetic process is the segmentation of the elongating body axis during which the precursors of the vertebrae are formed \cite{Oates2012}.
Segments form rhythmically and sequentially from an unsegmented progenitor tissue, the \emph{presomitic mesoderm} (PSM), see Fig.~\ref{fig.1}A.
During segmentation, the body axis elongates while the PSM continuously changes its length and eventually shortens.
After a species-dependent number of segments is formed, the process terminates when the PSM becomes very small \cite{Bellairs1986,Tenin2010}.
The total segment number can vary from about ten in frog to several hundreds in snake \cite{Gomez2008}.
The temporal progress of segmentation is controlled by oscillations of the cellular concentration levels of functional proteins in the PSM \cite{CookeZeeman1976,Oates2012}.
These cellular oscillations are achieved through autoregulation of so-called `cyclic genes' \cite{Lewis2003,Schroter2012,Webb2016}. On the tissue level, these oscillations give rise to nonlinear waves that propagate through the PSM \cite{Masamizu2006,Giudicelli2007,Aulehla2008,Morelli2009,Ares2012,Delaune2012,Lauschke2013,Soroldoni2014}, see Fig.~\ref{fig.1}B.
A new segment is formed with each completed oscillation at the anterior end of the PSM, corresponding to an arriving wave \cite{Soroldoni2014}.
Hence, in contrast to pattern formation via instabilities of homogeneous states, segmentation is characterized by a spatially inhomogeneous system with the PSM driving the patterning process.
In addition, pattern formation takes place in a dynamic medium: the body axis continuously elongates during segmentation while the PSM at the tail of the body axis shortens until segmentation terminates.
A theoretical model that integrates pattern formation, tissue elongation and shortening, and termination of segmentation by a self-organized mechanism is still missing.

To yield robust morphological results, this complex patterning process requires tight integration of spatial and temporal cues.
What regulates the integration of tissue growth and patterns of oscillating gene expression \emph{in vivo}?
The elongating body axis exhibits spatial concentration profiles of several signaling molecules.
In particular, the PSM displays posterior protein concentration profiles of FGF (fibroblast growth factor) and Wnt and an opposing anterior profile of Retinoic Acid \cite{Niederreither1997,Gomez2008,DiezdelCorral2003,Moreno2004,Aulehla2003,DobbsMcAuliffe2004,Dubrulle2004,Goldbeter2007,Shimozono2013,Bajard2014}, see Fig.~\ref{fig.1}C.
These signaling molecules are thought to be involved in body axis elongation \cite{Benazeraf2010} and regulating oscillations \cite{Ishimatsu2010} and cell fate during segmentation \cite{Aulehla2010}, i.e., maintaining cells in an oscillatory state within the PSM and triggering segment formation upon arrival at its anterior end.
Moreover, Retinoic Acid and FGF have been reported to display mutual inhibition and to antagonistically regulate genes that are involved in specifying segmented and unsegmented tissue \cite{DiezdelCorral2003,Moreno2004}.
Basic principles of segmentation were captured by simplified models in which the PSM length is kept constant \cite{Morelli2009,Ares2012,Murray2011}.
These studies use discrete or continuous phase descriptions of the cellular oscillators to describe the effects of cellular interactions on tissue level.
Furthermore, the role of signaling activity within the PSM  was studied using different approaches
including reaction-diffusion models of signaling gradients without coupling to oscillators or coupled to static clocks \cite{Baker2006a,Baker2006b,Baker2007}, models of signaling dynamics that is explicitly time-dependent \cite{Chisholm2011}, models based on diffusible cyclic gene products \cite{Cotterell2015} and detailed cell-based models \cite{Tiedemann2007,Hester2011,Tiedemann2012}.
However, whether the interplay of signaling activity and genetic oscillations alone can lead to self-organized gene expression waves and robust segmentation of the body axis in a dynamic tissue is not understood.

In this paper, we present a theoretical description of vertebrate segmentation, in which spatial concentration profiles of signaling activity regulate both growth and the frequency of cellular oscillators that control segmentation.
We show that local interaction rules can lead to robust self-organization of genetic oscillations, tissue elongation, and PSM shortening.
The resulting patterning system yields the correct spatial morphology and shortening of the segmenting tissue.
We introduce a one-dimensional continuum description of the dynamic tissue based on phase oscillators and two signaling activities $Q$ and $R$ varying in space and time, which represent Wnt/FGF and Retinoic Acid activity, respectively.
The signaling activities are effective tissue-level representations of the opposing and antagonizing signaling gradients found \emph{in vivo}, see Fig.~\ref{fig.1}C.
The interactions of the signaling system lead to termination of segmentation after a finite number of segments.
We study how the key features of segmentation depend on the kinetic parameters of the signaling system.
In a second step, using a simplified scenario with a single signaling gradient and time-periodic wave patterns, we derive analytical relations that explicitly show how the characteristic length scales of segments, waves, and tissue extension arise from our model.

\begin{figure}[t]
\begin{center}
\includegraphics[width=10.5cm]{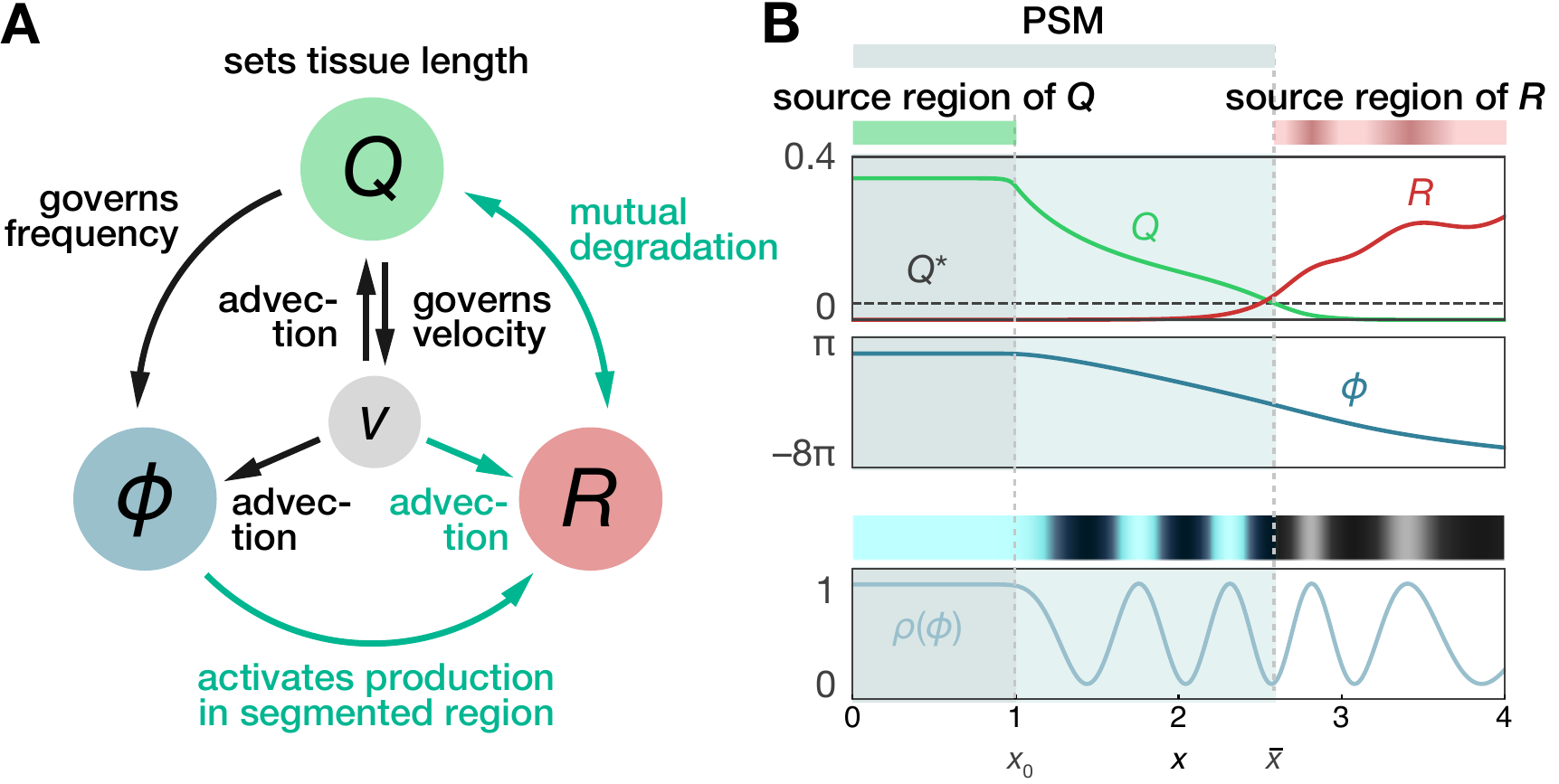}
\caption{(A) Interaction structure of the full model, Eqs.~(\ref{eq.phi}--\ref{eq.epsilon}). Interactions shown in green are not present in the reduced model given by Eqs.~(\ref{eq.phi.2}, \ref{eq.Q.2}). (B) Snapshot of a numerical solution to Eqs.~(\ref{eq.phi}--\ref{eq.epsilon}). The lower panel shows the associated wave pattern given by $\rho(\phi)=(1+\cos \phi)/2$. Parameters are $E=0.05$, $k=1$, $k'=150$, $\mu=1$, $D=1$, $h=5$, $h'=40$, $\nu=0.5$, $\nu'=3$, $Q^*=0.05$, $\kappa_0=0.15$, $\omega_0=4$, $\varepsilon_0=0.004$, $x_0=1$. The plots show the system at time $t=1$.}
\label{fig.2}
\end{center}
\end{figure}

We introduce a curved coordinate axis along the embryo, in which $x=0$ corresponds to the posterior tip of the PSM, see Fig.~\ref{fig.1}A.
We consider a set of cellular oscillators in the PSM, described by their phase $\phi$ in the oscillation cycle.
Growth of the tissue and frequency of the oscillators is regulated by a signal $Q$ that moves with the cell flow and is degraded, see Fig.~\ref{fig.2}A (black and gray arrows).
A second signal $R$ that emanates from the formed segments and quickly diffuses triggers additional degradation of $Q$.
The spatio-temporal distributions of the signals are described by the one-dimensional activity fields $Q(x,t)$ and $R(x,t)$.
Furthermore, we introduce a phase field $\phi(x,t)$ that represents the local state of the cellular oscillators in the PSM \cite{Morelli2009,Jorg2015}.
The dynamic equations for the phase field~$\phi$ and the concentrations $Q$ and $R$ are given by
\begin{eqnarray}
	\partial_t \phi + v \partial_x \phi &= \omega + \varepsilon \partial_x^2 \phi \ , \label{eq.phi} \\
	\partial_t Q + \partial_x (vQ) &= E \partial_x^2 Q - kQ-k'RQ  + \mu \sigma(x) \ , \label{eq.Q} \\
	\partial_t R + \partial_x (vR) &= D \partial_x^2 R	 - hR - h'QR + (\nu+\nu' \rho(\phi))\Theta(Q^*-Q) \label{eq.R} \ .
\end{eqnarray}
where $v(x,t)$ is a velocity field accounting for cell flow, $\omega(x,t)$ is the intrinsic frequency of the cellular oscillators, $\varepsilon(x,t)$ is the coupling strength. $D$ and $E$ are the diffusion constants of $Q$ and $R$, respectively, $\mu$ and $\nu$ are their basal production rates, $\nu'$ is the production rate of $R$ in the formed segments, $k$ and $h$ are decay rates. The rates $h'$ and $k'$ indicate the degree of mutual degradation of $Q$ and $R$, motivated by the antagonistic action of opposing gradients \emph{in vivo}. The dependence of the production rate of $R$ on $\rho(\phi)=(1+\cos \phi)/2$ leads to additional localized production in the center of the formed segments. For simplicity, we here consider a constant length $x_0$ of the source region of $Q$, $\sigma(x)=\Theta(x_0-x)$.
In our model, the local elongation rate $\partial_x v$ as well as the frequency and coupling strength profiles $\omega$ and $\varepsilon$ are controlled by $Q$ through the relations
\begin{eqnarray}
	\partial_x v &= \kappa_0 Q /Q^* \ , \qquad v|_{x=0}=0 \ , \label{eq.v} \\
	\omega &= \omega_0 Q/Q^* \ , \label{eq.omega} \\
	\varepsilon &= \varepsilon_0 Q/Q^* \ ,
	\label{eq.epsilon}
\end{eqnarray}
where $\kappa_0$, $\omega_0$, and $\varepsilon_0$ are a characteristic elongation rate, oscillation frequency, and coupling strength, respectively.
The position~$\bar x(t)$ of the anterior end of the PSM is defined as the point where the level of $Q$ reaches the threshold level~$Q^*$, 
\begin{eqnarray}
	Q(\bar x(t),t)=Q^* \ , \label{eq.psm.length}
\end{eqnarray}
see Fig.~\ref{fig.2}B. We consider open boundary conditions for the phase field, $\partial_x \phi |_{x=0} =0$.
For the signaling activities, we consider no-flux boundary conditions at the posterior tip, $\partial_x Q |_{x=0} = \partial_x R |_{x=0} =0$.

\begin{figure}[t]
\begin{center}
\includegraphics[width=10.5cm]{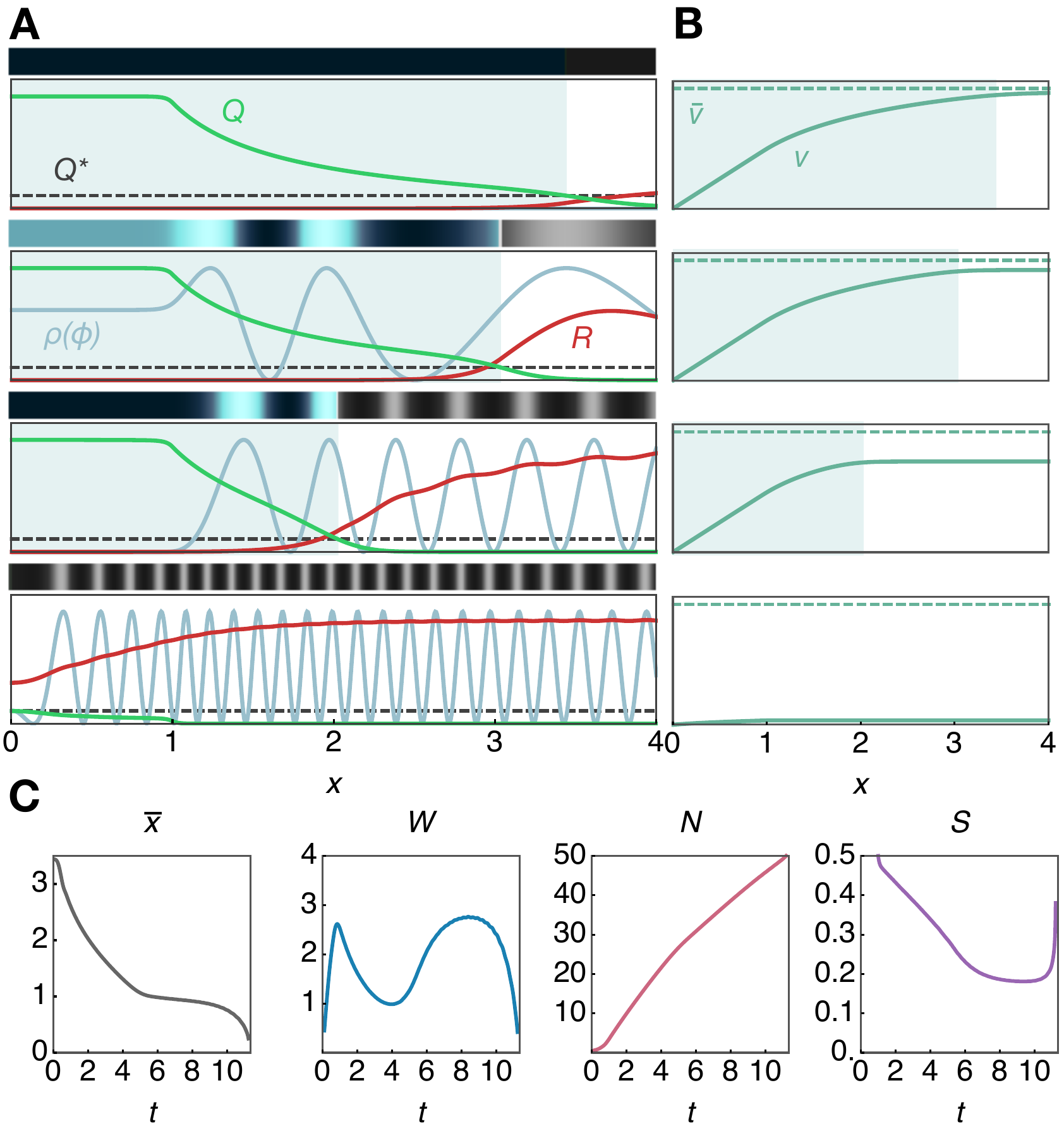}
\caption{(A) Snapshots of the time evolution of the system Eqs.~(\ref{eq.phi}--\ref{eq.R}) and (\ref{eq.v}--\ref{eq.epsilon}). Spatial distributions of $Q$ (green), $R$ (red) and $\rho(\phi)$ (blue) at different points in time (from top to bottom: $t=0,0.5,2,11.33$). The dashed line marks the threshold level $Q^*$ which also sets the $x$-axis scale for concentration levels. The colored top panel shows a density plot representation of $\rho(\phi)=(1+\cos \phi)/2$ showing the corresponding wave pattern.
(B) Velocity profile $v$ (solid curve) as determined from Eq.~(\ref{eq.v}) and $\bar v$ (dashed line) as velocity reference for the respective time points in A. In A and B, the shaded area marks the PSM region where $Q>Q^*$.
(C) Time evolution of the PSM length $\bar x$, defined through Eq.~(\ref{eq.psm.length}), the number of waves $W$, the segment number $N$, and the segment length $S$ at time of formation.
Parameters are the same as in Fig.~\ref{fig.2}.}
\label{fig.3}
\end{center}
\end{figure}

The model proposed here shows that the dynamics of segmentation and length decrease of the PSM can arise from local cellular interactions as a self-organized process.
To illustrate this, we start the system with a steady-state initial condition in which $Q$ and $R$ form opposing gradients in the absence of phase dynamics, see Fig.~\ref{fig.3}A.
The profile of $Q$ generates a spatial profile of intrinsic frequencies through Eq.~(\ref{eq.omega}).
In the source region, the profile of $Q$ is typically flat due to the balance of production, decay, and growth.
The frequency profile leads to a wave pattern, i.e., different parts of the PSM are out of phase \cite{Giudicelli2007,Morelli2009,Jorg2015}.
The pattern is flat in the posterior and displays a characteristic wavelength in the anterior, in accordance with experiments \cite{Soroldoni2014}.
This corresponds to no segments and a PSM of finite length with a flat phase profile.
Figs.~\ref{fig.3}A,B show snapshots of a numerical solution of Eqs.~(\ref{eq.phi}--\ref{eq.R}) with (\ref{eq.v}--\ref{eq.epsilon}) for different time points (see also Supplemental Movie 1).
As soon as waves leaves the PSM, i.e., as they enter the region where $Q<Q^*$, they become sources of $R$ according to Eq.~(\ref{eq.R}), see Fig.~\ref{fig.3}A.
The thus elevated levels of $R$ diffuse into the PSM and lead to an increased degradation of $Q$ in the vicinity of the anterior end (at $Q=Q^*$) and the profile of $Q$ shortens towards the posterior, see Figs.~\ref{fig.2}C and \ref{fig.3}A.
Consequently, the PSM shortens, see Fig.~\ref{fig.3}C.
Finally, the segmentation process terminates with the PSM reaching zero size, $\bar x=0$.

To discuss the dynamic features of this model, we define the time-dependent number of waves $W$, the segment length $S$, and the number of formed segments $N$ \cite{Jorg2015},
\begin{eqnarray}
	W(t) = \frac{\phi(0,t)-\phi(\bar x(t),t)}{2\pi} \ , \\[6pt]
	S(t) = \frac{2\pi}{|\partial_x \phi(\bar x(t),t)|} \ , \\[6pt]
	N(t)= \frac{\phi(\bar x(t),t)}{2\pi}
\end{eqnarray}
The number of waves is the total phase difference between posterior tip $x=0$ and anterior end $\bar x$, and the segment length is given by the local wavelength of the pattern at the anterior end.
Fig.~\ref{fig.3}C shows these functions together with the time evolution of the PSM length.
The time dependence of these quantities capture key features  of the segmentation process as seen in experiments, such as the decrease in PSM length, the formation time of segments, and the decrease of segment length over time \cite{Soroldoni2014,Jorg2015}.
The non-monotonic time dependence of the number of waves is not observed in experiments \cite{Soroldoni2014} and is related to the constant source length, which we have chosen here for simplicity.

\begin{figure}[t]
\begin{center}
\includegraphics[width=10.5cm]{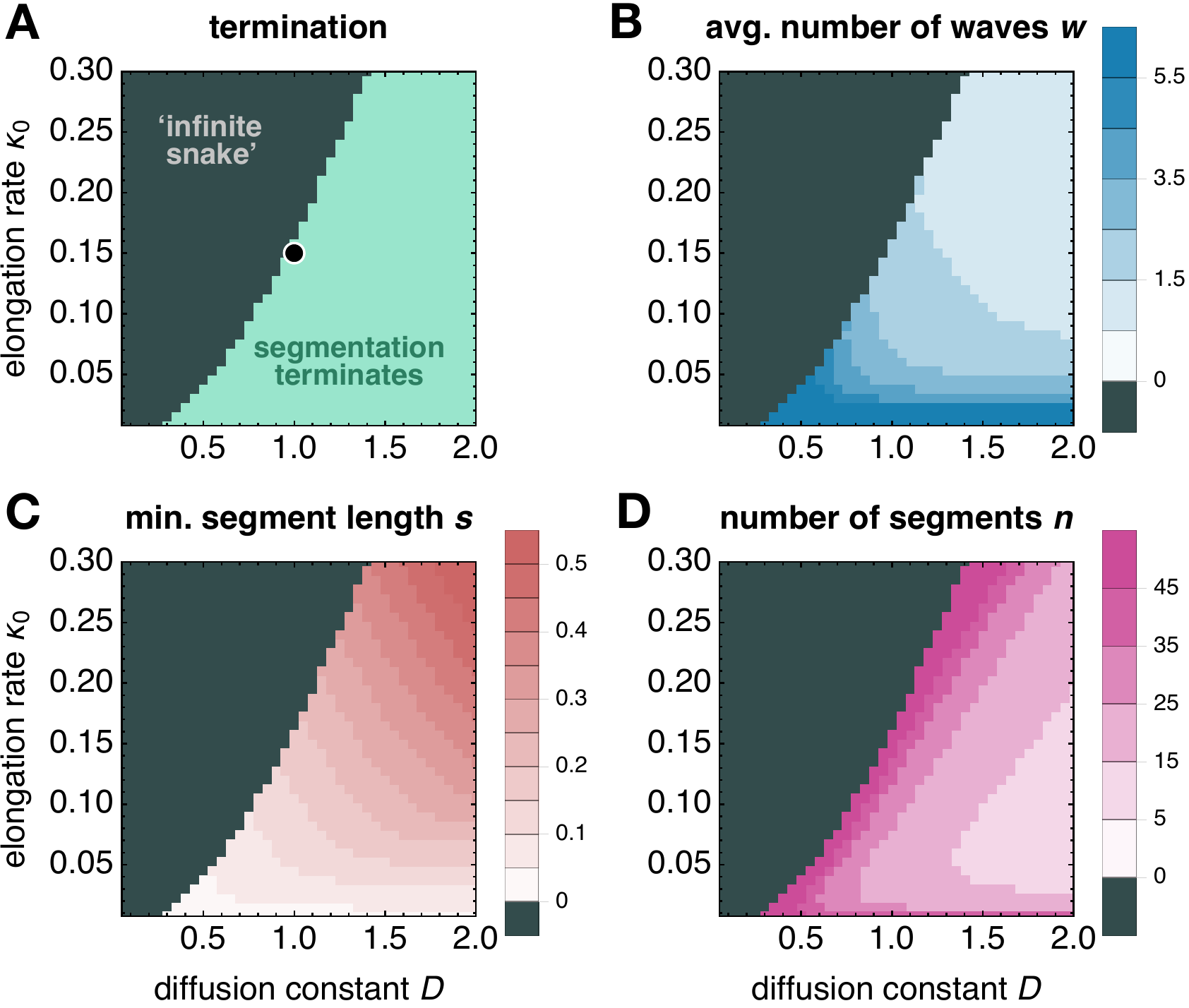}
\caption{
Key observables as a function of the diffusion constant $D$ of $R$ and the elongation rate $\kappa_0$.
(A) Regions in the parameter space in which segmentation terminates after a finite number of segments (green) and where the system attains a steady state at finite PSM length, forming infinitely many segments (gray).
The black dot marks the parameter set shown in Fig.~\ref{fig.3}.
(B) Average number of waves,
(C) total number of segments,
(D) minimum segment length.
The other parameters are as given in the caption of Fig.~\ref{fig.2}.
}
\label{fig.4}
\end{center}
\end{figure}

We now show that we can obtain a variety of different oscillation patterns and morphologies by changing the dynamics of signaling.
As an example, we vary two key parameters: the diffusion constant $D$ of the signaling component $R$, and the elongation rate $\kappa_0$, which sets the scale of cell flow velocity.
We define the total time $T$ of segmentation as the time at which the PSM has shortened to zero.
The average number of waves in the PSM is denoted by $w=\langle W(t) \rangle_{0\leq t \leq T}$, the minimal segment length is $s=\min_{0\leq t \leq T} S(t)$, and the total segment number is $n=N(T)$.
Fig.~\ref{fig.4} shows these observables as a function of $D$ and $\kappa_0$.
We find that the total time of segmentation diverges at the boundary shown in Fig.~\ref{fig.4}A.
Within a parameter range (gray), segments are generated in a time-periodic manner without end (`infinite snake') corresponding to the simplified theory Eqs.~(\ref{eq.phi.2},\ref{eq.Q.2}).
Biologically relevant parameters are found in the green region, where the total time $T$ of segmentation is finite.
This occurs if advection described by $\kappa_0$ is small enough that $R$ can diffuse sufficiently far into the PSM to degrade $Q$.
In this parameter region, the number of waves $w$ decreases with stronger advection, while the segment length $s$ increases, see Figs.~\ref{fig.4}B,C. The latter trend can be understood through the simplified model, see Eq.~(\ref{eq.S.sol}) and below.
Interestingly, the total segment number $n$ displays a non-monotonic behavior as a function of the elongation rate $\kappa_0$ with the smallest number of segments formed for intermediate values of $\kappa_0$, see Fig.~\ref{fig.4}D.

The model with two signaling gradients described by Eqs.~(\ref{eq.phi}--\ref{eq.epsilon}) can generate self-organized segmentation. However, additional insights in the length scales of segments, tissue, and wave pattern arise from the model dynamics can be obtained from the study of a simplified model.
In this reduced version of our model, only the posterior signaling gradient $Q$ is present, see Fig.~\ref{fig.2}A (black arrows only). This reduced model cannot account for PSM shortening but captures the self-organization of genetic oscillations in wave patterns and tissue elongation.
The dynamic equations for~$\phi$ and $Q$ are then given by
\begin{eqnarray}
	\partial_t \phi + v \partial_x \phi &= \omega + \varepsilon \partial_x^2 \phi \ , \label{eq.phi.2} \\
	\partial_t Q + \partial_x (vQ) &= - kQ  + \mu \sigma(x) \ . \label{eq.Q.2}
\end{eqnarray}
The system of Eqs.~(\ref{eq.phi.2}) and (\ref{eq.Q.2}) together with Eqs.~(\ref{eq.v}--\ref{eq.epsilon}) has a solution with a stationary activity profile $Q(x,t)=Q(x)$ and a time-periodic wave pattern $\phi(x,t)=\Omega t + \psi(x)$, where $\Omega$ is the collective frequency with which the wave pattern corresponding to the phase profile $\psi$ repeats.
Note that the dependence of $v$ on $Q$ makes Eq.~(\ref{eq.Q.2}) a nonlinear equation.
The number $W$ of waves in the PSM and the segment length $S$ at formation are now time-independent and given by 
\begin{eqnarray}
	W &= \frac{\psi(0)-\psi(\bar x)}{2\pi} \ , \label{eq.w} \\
	S &= \frac{2\pi}{|\psi'(\bar x)|} \ . \label{eq.s}
\end{eqnarray}

Interestingly, general properties can be discussed independent of knowledge of the full solution.
We here consider the case in which coupling only provides a minor correction to the phase pattern. As an approximation, we set $\varepsilon=0$ in Eq.~(\ref{eq.phi.2}).
For the phase profile $\psi$, we thus obtain \cite{Jorg2015}
\begin{eqnarray}
	\psi(x) = \int_0^x \frac{\omega(y)-\Omega}{v(y)} \, \mathrm{d}y \ . \label{eq.phase.profile}
\end{eqnarray}
The condition $v(0)=0$ determines the collective frequency as
\begin{eqnarray}
	\Omega = \omega(0) = \omega_0 Q_0/Q^* \label{eq.Omega.sol}
\end{eqnarray}
where $Q_0=Q(0)$.
Consequently, the segment length is given by $S=
(2\pi \bar v/ \omega_0) (Q_0/Q^*-1)^{-1}$, where $\bar v = v(\bar x)$.
The velocity $\bar v$ at the anterior end and the posterior concentration $Q_0$ can 
 be obtained from the stationary profile $Q(x)$, which, according to Eq.~(\ref{eq.Q.2}), satisfies
\begin{eqnarray}
	k Q = \mu \sigma(x) - \frac{\mathrm{d}}{\mathrm{d}x}(vQ) \ . \label{eq.Q.2.2}
\end{eqnarray}
Eliminating $Q$ in Eq.~(\ref{eq.Q.2.2}) by using Eq.~(\ref{eq.v}), we obtain an equation for the elongation rate $\mathrm{d} v/\mathrm{d}x$. Integrating from $0$ to $\bar x$ yields $\bar v =   \mu x_0 \lambda / Q^*$ for the case $\bar x \geq x_0$, where $\lambda=(1+k/\kappa_0)^{-1}$.
The posterior concentration $Q_0$ is determined by Eqs.~(\ref{eq.Q.2.2}) and (\ref{eq.v}) at the posterior boundary $x=0$ as ${Q_0}/{Q^*}=[(k^2+4\kappa_0 \mu/Q^*)^{1/2}-k ]/2\kappa_0$. 
Hence, the segment length is given by
\begin{eqnarray}
	S &= \frac{\mu \lambda T_0}{Q_0-Q^*} x_0 \ , \label{eq.S.sol}
\end{eqnarray}
where $T_0=2\pi/\omega_0$. Since $\mathrm{d} \lambda/\mathrm{d} \kappa_0>0$ and $\mathrm{d} Q_0/\mathrm{d} \kappa_0<0$, the segment length $S$ increases with increasing elongation rate.
Note that $S$ is proportional to the length $x_0$ of the source region, which is the only length scale in the system.

The full solution for the steady state of $Q$ is given by
\begin{eqnarray}
	Q = Q_0 \times \cases{1 &for $x < x_0$\\
	\frac{1-\beta}{1+\mathcal{W}(-\beta^{-1} \mathrm{e}^{-\beta^{-1}(1+(1-\beta)^2 (x-x_0)/x_0)} )^{-1}} &for $x>x_0$\\} \label{Q.steadystate}
\end{eqnarray}
where $\beta = 1+(k/\kappa_0) (Q^*/Q_0)$ and $\mathcal{W}$ is the principal branch of the Lambert $\mathcal{W}$ function, defined by the relation $\mathcal{W}(z) \mathrm{e}^{\mathcal{W}(z)}=z$ \cite{Corless1996}.
The growth field generated by $Q$ through Eq.~(\ref{eq.v}) corresponds to a velocity field, where cells move anteriorly and reach their maximum speed at the anterior end of the tissue.
From Eq.~(\ref{Q.steadystate}), the PSM length $\bar x$ can be obtained using the definition Eq.~(\ref{eq.psm.length}),
\begin{eqnarray}
	\bar x =  \frac{\beta}{(1-\beta)^2} (\beta+\lambda-\log \beta \lambda-2) x_0 \ . \label{eq.xbar.sol}
\end{eqnarray}
The number of waves $W$ can be obtained using Eqs.~(\ref{eq.phase.profile}) and (\ref{eq.omega}) and approximating the velocity field by the velocity at the anterior end, $v(x) \simeq \bar v$,
\begin{eqnarray}
	W 
	&\simeq 
\frac{\Omega}{2\pi} \left( \frac{\bar x}{\bar v} + \frac{1-\beta}{k} \right) \ . \label{eq.W.sol}
\end{eqnarray}
Dynamic solutions of the minimal model with constant PSM length are shown in Supplemental Movie 2.
The stationary profiles for $Q(x)$ and $\psi(x)$ describe the periodic wave-like pattern of gene expression that is determined by a signaling gradient.
These solutions are similar to the stationary `infinite snake' patterns found in the full model, see Fig.~\ref{fig.4}A.
The waves propagate in a dynamic tissue that expands with a velocity profile determined by Eq.~(\ref{eq.v}).

In this paper, we have shown how segmentation of the vertebrate body plan can arise as a self-organized patterning process controlled by signaling activity coupled to genetic oscillators.
In our full model, local rules representing cellular interactions lead to (i) self-organized wave patterns, (ii) tissue elongation, (iii) PSM shortening, and (iv) self-organized termination of segmentation with a finite number of segments.
To complement our study, we used a reduced model with one signaling gradient to derive explicit relations between the time and length scales of the wave pattern and segmentation and the biochemical properties of signaling.
Our work shows that by varying biochemical parameters of the signaling gradients, a variety of different patterns and morphologies can be generated.
Such variations could correspond to the differences in the segmentation process and the resulting morphologies between different species, e.g., in fish, mouse, chick, snake, and frog.
Among these species, the observed number of segments ranges from about ten to several hundred, the number of waves ranges from one to five \cite{Gomez2008}.
In our model, the length of the signaling source region sets the length scales of wave patterns and segment length. This correspondence could naturally account for `scaling' of segment size with the size of the organism if this source region occupies a characteristic proportion of the segmenting tissue.
Note that the principle of self-organization proposed here does not require diffusion of the posterior signaling molecule in the tissue.
Indeed, our work shows that frequency and growth profiles could emerge either from effective diffusion or from advection or from a combination of both \cite{Lecuit1996,Chisholm2010}.
It is worth noting that our model also exhibits two recently discovered wave effects in embryonic segmentation: a Doppler effect that arises from the anterior end of the PSM moving into the waves and a `dynamic wavelength effect' that denotes the decrease of the wavelength at a fixed position over time \cite{Soroldoni2014} (both effects can be inspected in Supplementary Movie 1).

\begin{figure}[t]
\begin{center}
\includegraphics[width=10.5cm]{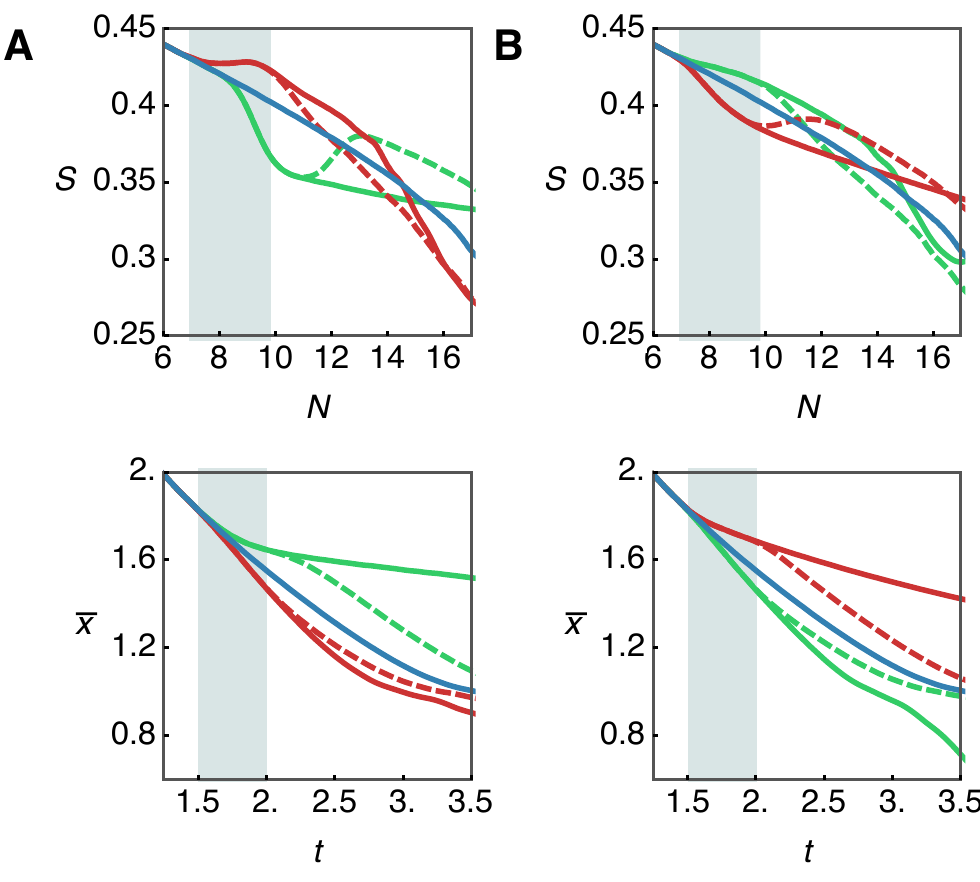}
\caption{
Effects of perturbations of signaling gradients. Segment length $S$ as a function of segment number $N$ and PSM length $\bar x$ as a function of time $t$ for the system displayed in Fig.~\ref{fig.2} but with perturbations of (A) the signal $Q$ and (B) the signal $R$. The curves show upregulation (green), downregulation (red), and the unperturbed system (blue) for reference. Solid curves indicate a sustained perturbation starting from $t=1.5$, dashed curves indicate a transient perturbation from $t=1.5$ to $t=2$ (gray area). Perturbations were realized by increasing production rates ($\mu$, $\nu'$) or degradation rates ($k$, $h$) according to $\mu \to 1.3\mu$ (green), $k \to 1.3k$ (red) in panels (A), $\nu' \to 1.3\nu'$ (green), $h\to 1.5 h$ (red) in panels (B).
}
\label{fig.5}
\end{center}
\end{figure}

There are several possibilities to test whether the mechanism proposed here actually underlies the segmentation process \emph{in vivo}.
One of these possibilities is transient up- or downregulation of the signaling molecules present in the PSM.
Experimentally controlled transient reduction of Wnt signaling, for instance, leads to a shorter average PSM length, faster PSM shortening, and longer segments \cite{Bajard2014}, all observations consistent with our model, see Fig.~\ref{fig.5}A (red curves).
Furthermore, our model could be tested by experimentally induced transient up- or downregulation of Retinoic Acid in a quantitative and dynamic assay system, see Fig.~\ref{fig.5}B.
The details of cross-regulation between signaling molecules are yet largely unknown and how, for instance, signaling activity regulates the frequency of the cellular oscillators and cell fate on a molecular level remains an open question.
Our model, which is based on available experimental knowledge, yields testable predictions on tissue level that could shed light on the nature of these molecular interactions and motivate further experimental and theoretical research.

\section*{Acknowledgments}
We thank Rachna Narayanan and Luis Morelli for fruitful discussions.
This work was supported by the Francis Crick Institute which receives its core funding from Cancer Research UK, the UK Medical Research Council, and the Wellcome Trust. In addition, AO was supported by a Senior Research Fellowship from the Wellcome Trust (WT098025MA) and the Medical Research Council (MC\_UP\_1202/3). \\

\providecommand{\newblock}{}

\end{document}